\documentclass[mathleft
]{an}
\usepackage{graphicx}
\usepackage{times}
\begin{document}

\Pagespan{789}{}
\Yearpublication{2006}%
\Yearsubmission{2012}%
\Month{01}%
\Volume{999}%
\Issue{88}%

\title{Space Astronomy for the mid-21st Century: \\  Robotically Maintained Space Telescopes}

\author{N. Schartel\inst{1}\fnmsep\thanks{Corresponding author:
  \email{Norbert.Schartel@sciops.esa.int}\newline}
}
\titlerunning{Robotically Maintained Space Telescopes}
\authorrunning{N. Schartel}
\institute{
ESA, ESAC, XMM-Newton Science Operations Centre, 
Apartado 78, 28691 Villanueva de la Ca{\~nada}, Spain}

\received{10 January 2012}
\accepted{14 February 2012}
\publonline{? ? 2012}


\abstract{%
The historical development of ground based astronomical telescopes leads us to expect that space-based astronomical telescopes will need to be operational for many decades. 
The exchange of scientific instruments in space will be a prerequisite for the long lasting scientific success of such missions. 
Operationally, the possibility to repair or replace key spacecraft components in space will be mandatory. 
We argue that these requirements can be fulfilled with robotic missions and see the development of the required engineering as the main challenge. 
Ground based operations, scientifically and technically, will require a low operational budget of the running costs. 
These can be achieved through enhanced autonomy of the spacecraft and mission independent concepts for the support of the software.      
This concept can be applied to areas where the mirror capabilities do not constrain the lifetime of the mission. 
}

\maketitle

\section{Introduction}

Today, about 20\% of the articles in major astronomical journals (Astronomy and Astrophysics, The Astrophysical Journal and Monthly Notices of the Royal Astronomical Society),  refer to X-rays in their title or abstract. 
This illustrates the fact that X-ray observations have become one of our main tools of astrophysical research. 
This development is based on a fleet of satellites with three observatory-type facilities 
 currently operating: $\;$ Chandra (Weisskopf et al. 2002), XMM-Newton (\cite{Jansen2001}) and Suzaku (\cite{Mitsuda2007}). 
During the last thirty years the astronomical community had almost permanent access to X-ray 
 observations through a multitude of national and international missions, 
 e.g. HEAO-1, Einstein, EXOSAT, Ginga, ROSAT, ASCA and  BeppoSax.

Active astronomers expect to have permanent access to the X-ray sky with observatory-class facilities $\;$ to satisfy their scientific needs. 
Although the funding agencies recognize this demand as demonstrated by the high priority that the next generation X-ray telescopes
received in US Decadal Survey and in ESA's Cosmic Vision, the future is far from certain due to budget constraints and the
requirement to explore other observational areas $\;$ through space born facilities, e.g. the gravitational waves. 
Even if there would be an approved next-generation X-ray telescope, the basic expectation, permanent access to the X-ray sky would only be temporarily be satisfied due to the restricted  lifetime of space missions.

In this paper we try to explore if ground based astronomy and its recent history can be a helpful guide. 
In section~\ref{p2}  we outline basic development in  ground based astronomy and we present in section~\ref{p3} a concept for space mission. 
In section~\ref{p4}  we discuss the  technical and organizational feasibility followed by some concluding remarks in section~\ref{p5}.

\section{Ground observatories}
\label{p2}

Over the last century ground based astronomical observing facilities were the subject to a major developmental process. 
Today, first class observatories are generally funded either nationally or even multi-nationally, e.g. ESO, ALMA, etc. 
The primary characteristic of a telescope is its collecting area, and substantial scientific impact can be expected if it is possible to increase the collecting area by factor of ten (greater than five as minimum). 
The development cycle for such mirror advances is some thirty years. 
Instrumentations develop at a much faster rate. Within some ten years, instrumentation shows impressive progress.  
Construction of a telescope mounting, a dome and the mirror itself are the major costs. 
In comparison to these costs, the instrumentation requires only minor resources. 
Operations are an important aspect because their costs have to be paid annually. 
Generally, an observatory supports several telescopes allowing the minimization of the operational costs due to a high amount of synergy between operations of the individual telescopes. 
A leading observatory has to provide a new primary telescope every thirty years in order to continue to be scientifically competitive, whereas the instrumentation is replaced on shorter time scales. 
In general, older telescopes continue to be operated reaching operational lifetimes of some sixty years maximizing the scientific return on the primary investment.

\section{Outline of a concept}
\label{p3}

If we want to transfer concepts of ground based observatories into space then we must consider space borne telescopes with operational lifetimes of forty to sixty years. 
It is obvious that such telescopes will need maintenance in space. 
Only robotic maintenance missions can be expected to provide these at reasonable costs. 
Beside repair and replacement of failing components, and substitution of consumables, they will transport and install new
instruments such that the scientific value of the observatory $\,$ remains competitive. 
The robotically maintained space telescope $\,$ should include the following components: 
\begin{enumerate}
\item
 Selection of roboticaly maintained space telescope: 
 It is impossible to predict the astronomical requirements or the development of scientific instruments on timescales of thirty to sixty years. 
Therefore, the main criteria for the decision to build a robotic maintained space telescope must be the progress achieved in the engineering, construction and manufacturing of the satellite and the mirror and specifically the possibility to increase of collecting area of the mirror. 
We can assume that an increase in the effective area by a factor of 10 ($>$5) will bring substantial scientific progress and justify the selection of a mission. 
A further important criterion is the achievable sensitivity in comparison with other wavelengths $\,$ as the scientific merit $\,$ increases with wider wavelength coverage. 
Depending on the wavelength, an increase of the spatial resolution or other performance parameter may be an important addition.
\item
The space telescope: 
The primary goal is a long lifetime, i.e. a designed lifetime of some thirty years with a potential to operate for sixty years. 
Because it can not be expected that a mission will operate for such a long time without any failure, the mission must be constructed for robotic maintenance in space. 
Most important will be the possibility to replace consumables and the possibility to replace key spacecraft components, e.g. gyroscopes or reaction wheels. 
From the scientific side it is most important that the spacecraft allows a simple exchange of the scientific instruments.   
\item
Robotic maintenance missions: 
Assuming a typical instrumental development cycle $\,$ of ten years, $\,$ robotic maintenance missions $\,$ may visit the space telescope once each decade. 
The maintenance spacecraft must be able to perform small repairs, substitute key spacecraft elements, refill consumables and specifically to replace scientific instruments. 
The replacement of the scientific instrument is of utmost importance to ensure the scientific competitiveness of the mission. 
\item
 Operations:
 Given the long lifetime of the space telescope it is of importance that the ground operations run with extremely low costs. 
These can be ensured by a highly automated $\,$ and autonomous telescope and onboard instruments. 
Another aspect is the software used for ground operations. 
The challenge here is that the software has to be useable over many decades.  
A further important issue is the amount of funds the community can contribute to the scientific aspects of the operation of the mission, e.g. instrument calibration etc..
\end{enumerate}

\begin{figure}[thp]
\label{figure1}
\begin{center}
\includegraphics[width=57mm,angle=-90]{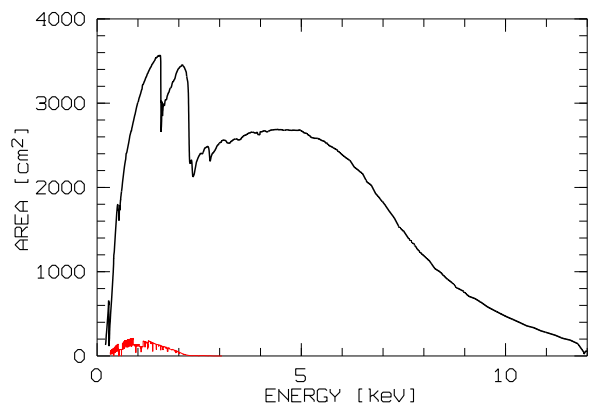}
\end{center}
\caption{
This figure illustrates the huge potential benefit of replacing the scientific instruments on XMM-Newton with microcalorimeter spectrometers. 
The black curve shows the effective area of hypothetical X-ray microcalorimeter spectrometers behind the three mirror modules on board of XMM-Newton. 
Details of the calculations are provided in the appendix \ref{p6}. 
For comparison the effective areas of the two currently operating Reflection Grating Spectrometers are given in red where we added the effective area of both spectrometers and the 1$^{st}$ and 2$^{nd}$ order. }
\end{figure}

\section{Feasibility of a robotically maintained space telescope} 
\label{p4}

Almost no challenge for robotically maintained space telescopes is new, unsolved or not already a hot topic of development and engineering and therefore is expected to be available within a decade. 
In the following section we give some comments on the arguments brought forward above and provide examples to illustrate the claims:
\begin{enumerate}
\item
 Selection of $\,$ roboticaly maintained space telescope: $\,$
\cite{Lang2010} illustrated impressively that astronomical research was and is fundamentally driven from unexpected detections which were possible only due to technological advance. 
The thirty years long research $\,$ to identify the nature of gamma ray bursts (\cite{Klebesadel1973} and \cite{vanParadijs1997}), illustrates the importance to have comparable sensitivity at different wavelengths combined with adequate spatial resolution. 
A good example is the current debate on the electro-magnetic $\,$ counterparts of gravitational wave sources. 
Specifically $\,$ the X-ray energies are considered most promising \cite{Komossa2010}, implying that a gravitat--ional wave telescope should be operated simultaneously with a large X-ray telescope. 
Only when we achieve simultaneous coverage at different wavelengths will the astronomical progress be optimal and efficient. 
\item
The space telescope: $\,$
 With respect to the space telescope, the foremost question is the scientific competitiveness of the collecting area on timescales of thirty to sixty years.
Fig.~1 demonstrates that the effective area of hypothetical microcalorimeter spectrometers on board of  XMM-Newton  would exceed the effective area of the currently operating Reflection Grating Spectrometers (RGS, \cite{denHerder2001}) by more than a factor ten. 
Both instruments can reasonably be assumed to have comparable spectral resolution.
Currently, there is no plan for a mission that has spatial resolution comparable to Chandra. 
Therefore, there will be a continual scientific demand that can only be satisfied by Chandra observations and there is not successor likely be to happen for some decades. 
A further example is ROSAT \cite{Truemper1982}, which operated from 1990 to 1999. 
Its grasp in the energy range from 0.1 to 2.0 keV was significantly higher that the grasp of XMM-Newton. 
In combination with the low background due to ROSATs orbit, large surveys aiming for cosmological studies could be performed with an operating ROSAT in about half the time required by XMM-Newton. 
These three examples illustrate that a conservation of telescop collecting areas for some thirty years is  scientifically compelling. 
Chandra and XMM-Newton are now operating for twelve years \cite{Santos-Leo2009}.
IUE reached an operational lifetime of eighteen years  \cite{Wamsteker2000} and HST is now twenty-two years old \cite{Dalcanton2009}. 
Therefore, lifetimes of sixty years are within the range of technological possibilities. 
HST is the only astronomical satellite which was constructed for maintenance in space and several components and instruments were replaced by astronauts. 
Most probably robotically maintained space telescopes will require a modular construction of the spacecraft $\,$ allowing easy replacements of complete units and therefore reducing the complexity of the robotic maintenance missions. 
Such modular elements may offer the possibility of general cost reduction because the modules might be used by several robotically maintained scientific and commercial satellites. 
The planning for the replacement of scientific instruments with maintenance missions significantly reduces the risk in the development of the primary generation of instruments that can constraint to established technology. 
\item
Robotic maintenance missions: 
The engineering of the spacecraft for robotic maintenance is the most challenging task for this concept. 
To development the required engineering skills for a single specific mission or task would most probably be inefficient due to the high costs.  
Robotic spacecrafts must maintain a wide range of commercial and scientific missions and may either be reused or built in large numbers. 
However, international, intergovernmental, national as well as private companies are currently developing such spacecrafts and the main issue is to utilize these developments also for scientific purposes. 
Examples for developments in this direction are the Special Purpose Dexterous Manipulator of 
 MacDonald, Dettwiler and Associates (MDA), the Robotic Refueling Mission of NASA, the \$280 million contract between Intelsat and MDA to develop robots for simple repairs and refueling of geostationary communication satellites and the Deutsche Orbital Servicing Mission of DLR for repair and non-destructive capturing of satellites. 
It is important to see that such robotic maintenance mission will have very low costs, e.g.  a refueling mission is estimated to cost some thirty to fifty million dollars.  
With respect to the exchange of scientific instruments it is important that scientific instruments generally have very low mass in comparison to the weight of the spacecraft and its mirror. 
ROSAT had a mass of some 2,500 kg, XMM-Newton has a mass of 3,800 kg and Chandra has a mass of 4,800 kg, respectively. 
For comparison a X-ray microcalorimeter spectrometer with cryogenic cooling has a mass of some 400 kg and a polarimeter has a mass of some 10 kg. 
The overall risk in association with the development and employment of new instruments is significantly reduced because the robotic $\,$
maintenance mission $\,$ would be easier $\,$ and cheaper than the primary mission and can be delayed. 
\item
Operations:
 Operational costs are important due to the long lifetime of the mission. 
Ideally astronomers should be able to operate $\,$ the satellite $\,$ within their research work. 
There are strong development lines in this direction. $\,$
Now, many astronomical satellites $\,$ are contin--uously operating for days without interaction with the ground station. 
The Swift satellite already allows astronomers to change the pointing direction with a cell phone call. 
Instruments can be developed such that no pointing direction will put them on risk and the instruments can be able to cope
 autonomously with unexpected source fluxes $\,$ or radiation environment.  $\,$
The maintenance of operational and scientific software over decades is challenging. $\,$
The solution can only be to avoid mission specific software as far as possible and be in a modular system structure. 
Many examples of task specific software are already applied in astrophysical and operational contexts.  
Ideally the individual missions would define their specific requirements simply in the form of configuration files whereas the software itself is unchanged. 
Ideally, astronomers could easily install the software by themselves. 
The HEASARC remote proposal submission system is an example for such a task specific software component. 
Scientific analysis software packages, like MIDAS or IRAF, are further examples of widely used mission independent software. 
Although the investments to establish a full set of operations and scientific software along the engineering concepts is high, the pay off would be threefold: (a) astronomers would not have to spend time to familiarize themselves with new software, (b) the costs of further science operation center development would be drastically reduced, and (c) the development risk would basically be eliminated.
\end{enumerate}

\section{Conclusions}
\label{p5}

Most of the examples provided come from X-ray astronomy and reflect experiences of the author. 
However, the concept of robotically maintained space observatories itself are naturally wavelength independent. 
The concept may be applied at any wavelength where the mirror capacities are sufficiently advanced such that the number of potential targets does not naturally constrain the lifetime of the mission. 
The application of the concept further depends on the ratio of instrument mass to space telescope mass and the development cycle of the instruments. 
The concept depends on the progress in robotic engineering, but currently enormous efforts and progress are being made. 
The scientific community is asked to consider and utilize the progress made in robotics for their future space missions.

\acknowledgements
I thank Brian McBreen, who convinced me to formulate my ideas of robotically maintained space telescopes 
 and to share them with the community. 
I would like to thank R. Gonz{\'a}lez-Riestra for calculating the RGS effective area 
 and M. Smith for his help with the XMM-Newton filter transmission.

\newpage


\appendix

\section{Effective area of hypothetical X-ray microcalorimeter spectrometer onboard of XMM-Newton}
\label{p6}

XMM-Newton has three X-ray telescopes. 
To estimate the total effective area of hypothetical X-ray microcalorimeter spectrometers 
 onboard XMM-Newton we considered three components: 
 the effective area of the telescopes, 
 the filter transmission and 
 the quantum efficiency of the microcalorimeter. 

The pn-CCD camera \cite{Strueder2001} is operated in the focal plane of X-ray telescope 3 
 which is the telescope without reflection grating array unit. 
We use the on-axis effective area of telescope 3 as provided in the current calibration file 
 XRT3\_XAREAE F\_0011.CCF\footnote{http://xmm2.esac.esa.int/external/xmm\_sw\_cal/calib/rel\_notes/index. shtml}. 
We approximate the transmission of the optical filters of the microcalorimeter through 
 the transmission of the thick filter of pn (EPN\_FILTERTRANSX\_0014.CCF$^1$). 
We assume a quantum efficiency (including filling factor and 'dead' detector elements) of 0.93 
 for energies below 4.5 keV which decreases with $\sim$0.11~keV$^{-1}$ for higher energies. 
The obtained total effective area is added as on line material (xmm\_cal.fits).

\end{document}